\documentclass{article}
\usepackage{spconf,amsmath,graphicx,hyperref}
\usepackage{booktabs}

\title{InstCharVoice: Grounding Natural-Language Instructions for Character-Level Control in Text-to-Speech}
\name{\shortstack{
    Sihang Nie$^{1}$\sthanks{Work conducted when the author was intern at Huya Inc.}, 
    Xueru Li$^{1*}$,
    Xiaofen Xing$^{1}$\sthanks{Corresponding author.}, 
    Deyi Tuo$^{2}$,
    Cheng-Bin Jin$^{2}$,
    Jingyuan Xing$^{1}$,
    Jinxin Ji$^{3}$
}}
\address{
  $^{1}$South China University of Technology,
  $^{2}$Huya Inc.,
  $^{3}$The Hongkong Polytechnic University
}

\begin{document}
\ninept
\maketitle
\begin{abstract}
Instruction-based text-to-speech (ITTS) systems enable natural-language control of expressive speech generation, but often offer limited transparency and fine-grained control over individual text units. Character-level controllable TTS systems provide explicit acoustic control, yet typically rely on user-specified acoustic attributes. To bridge this gap, we propose InstCharVoice, a unified framework that grounds natural-language instructions in character-level acoustic control. We first construct grounded instruction annotations on the WordVoice-5A-zh corpus using Qwen3-Omni. With this supervision, we train an autoregressive model to identify instruction-relevant characters and predict their acoustic attributes before generating the corresponding speech tokens. Keyword prediction and grounding-aware loss weighting help the model focus on instruction-relevant characters and attributes. Experiments show improved instruction following and keyword-level acoustic control over representative ITTS systems, with competitive speech naturalness and explicit character-level controllability. Audio samples are available at~\url{https://xxh333.github.io/instcharvoice-demo/}.
\end{abstract}
\section{Introduction}
\label{sec:intro}

Neural text-to-speech (TTS) systems have achieved impressive expressiveness in zero-shot voice cloning. Benefiting from large-scale speech data and generative modeling, these systems can synthesize speech with high naturalness and speaker similarity~\cite{cosyvoice2, f5tts, saras, qwenaudio3tts}. This progress has increasingly shifted research attention toward expressive instruction-based speech generation~\cite{emovoice, hdppt, ovinstructtts, voicesculptor, fireredtts3}.

Despite this progress, existing instruction-based TTS (ITTS) systems often provide limited transparency and fine-grained control. Many systems directly generate speech from text and instructions without explicitly identifying which text units should be modified or which acoustic attributes should realize the requested changes~\cite{ovinstructtts,question1}. Consequently, the correspondence among instruction phrases, target text units, and acoustic variations remains implicit. Moreover, existing ITTS systems typically emphasize utterance-level or segment-level characteristics, leaving precise control over individual characters less explored~\cite{voxinstruct,wordvoice}.

Fine-grained controllable TTS offers a complementary solution by representing local acoustic variations with explicit character-level attributes~\cite{wordvoice,magictts}. Such representations enable precise and inspectable control over individual text units. However, these methods typically rely on user-specified acoustic attributes, which can become cumbersome to configure as the utterance length and the number of controllable attributes increase. This attribute-based interface is less intuitive than expressing the desired changes in natural language. An important gap therefore remains between natural-language instruction control and explicit character-level acoustic realization.

\begin{figure}[t!]
\centering
\includegraphics[width=\linewidth]{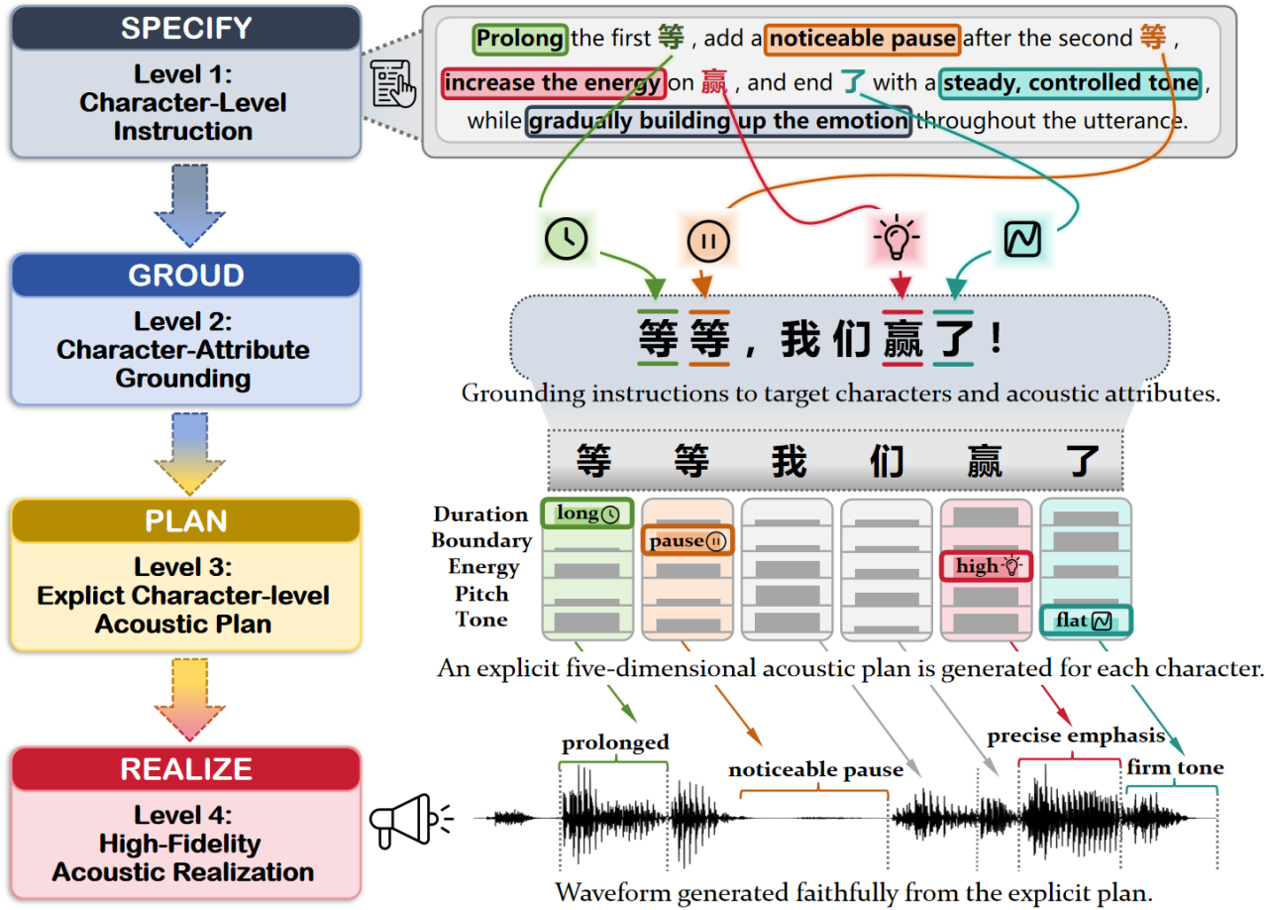}
\caption{Motivation and overview of InstCharVoice. The framework bridges natural-language instructions and explicit character-level acoustic control through instruction grounding and acoustic planning.}
\label{fig:mot}
\end{figure}

To bridge this gap, we propose \textbf{InstCharVoice}, an ITTS framework for fine-grained character-level control. As illustrated in Fig.~\ref{fig:mot}, InstCharVoice connects instruction specification, character-level grounding, acoustic planning, and speech realization. It identifies instruction-relevant characters and predicts five acoustic attributes---duration, boundary, energy, pitch, and tone---for each character before generating its corresponding speech tokens. To provide supervision for this grounded modeling process, we develop an LLM-assisted annotation pipeline that leverages an open-source multimodal language model to generate natural-language instructions and character-level grounding labels from text, speech, and acoustic attributes.

\begin{figure*}[htbp]
    \centering
    \includegraphics[width=0.97\linewidth]{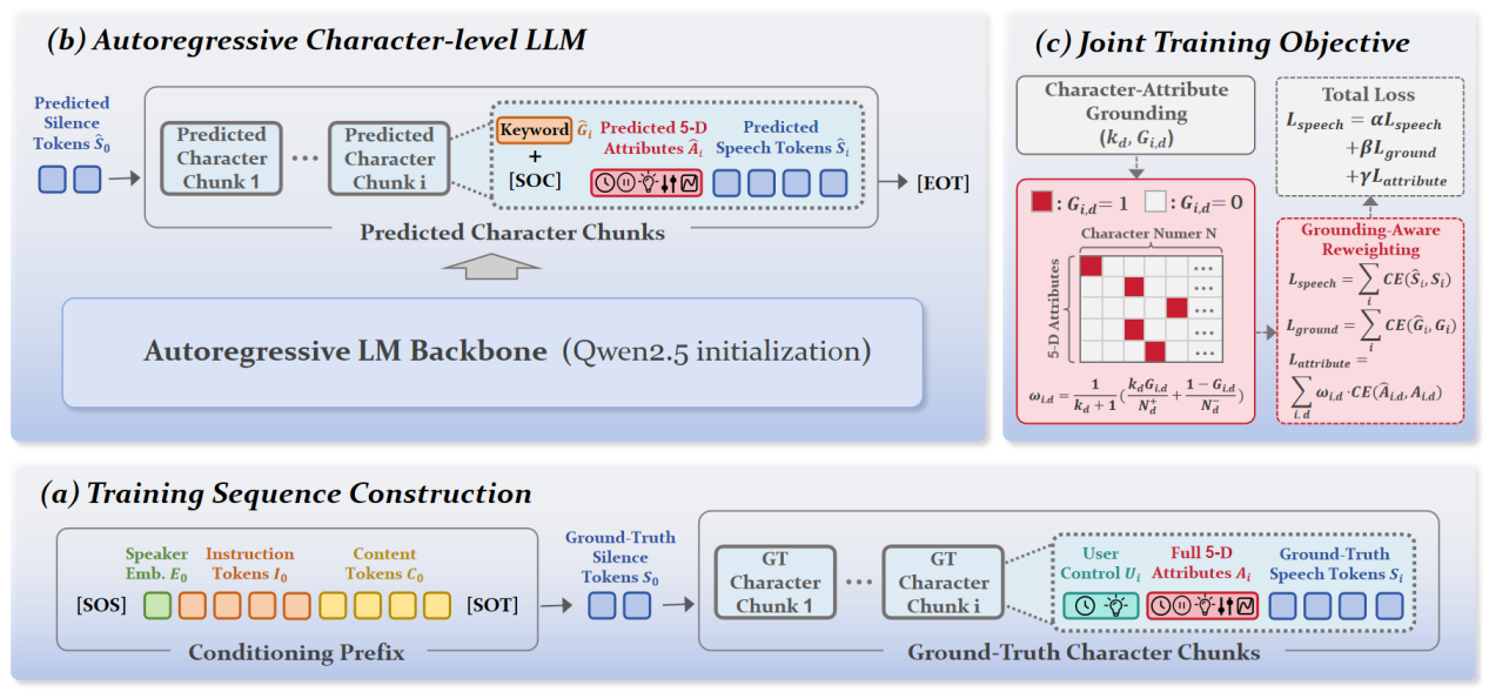} 
    \caption{Model architecture of InstCharVoice, including (a) training sequence construction, (b) autoregressive modeling of character-level speech chunks, and (c) grounding-aware training objective.}
    \label{fig:llm}
\end{figure*}

Our contributions are summarized as follows.
1) We develop an LLM-assisted annotation pipeline that links natural-language instructions to target characters and acoustic attributes, providing explicit supervision for instruction grounding.
2) We propose \textbf{InstCharVoice}, a unified framework that bridges natural-language instructions and speech generation through character-level acoustic planning, with keyword prediction and grounding-aware loss weighting to strengthen localized control.
3) Experiments show improved instruction following and keyword-level acoustic control over representative ITTS systems, while maintaining competitive speech naturalness and character-level controllability.

\begin{table}[t]
\caption{Statistics of the character-level grounded instruction annotations. Dur, Bnd, Eng, Pit, and Ton denote duration, boundary, energy, pitch, and tone, respectively.}
\label{tab:annotation_statistics}
\centering
\resizebox{\linewidth}{!}{%
\begin{tabular}{@{}l|rrrrrrr@{}}
\toprule
\textbf{Statistic} & \textbf{Utterances} & \textbf{Keywords} & \textbf{Dur} & \textbf{Bnd} & \textbf{Eng} & \textbf{Pit} & \textbf{Ton} \\
\midrule
\textbf{Count}
& 2143k & 5978k & 609k & 3882k & 2830k & 202k & 901k \\
\bottomrule
\end{tabular}
}
\end{table}

\section{Methodology}
\label{sec:method}

\subsection{Character-Level Grounded Instruction Annotation}
\label{sec:annotation}

We construct character-level grounded instruction annotations on WordVoice-5A-zh~\cite{wordvoice}, a corpus derived from LEMAS~\cite{lemas}. The dataset contains approximately 2,546 hours of Chinese speech, with transcriptions and character-level annotations for duration, energy, pitch, boundary, and tone. We develop a three-stage annotation pipeline based on Qwen3-Omni-30B-A3B~\cite{qwen3omni}~\footnote{https://www.modelscope.cn/models/Qwen/Qwen3-Omni-30B-A3B-Instruct}.

First, Qwen3-Omni generates natural-language instructions from each speech-text pair to describe utterance-level expressive and acoustic characteristics. 
We then provide the character-level acoustic attributes as explicit cues to refine the descriptions and correct inconsistencies. 
Finally, salient local acoustic patterns, such as prolonged duration or prominent energy, are extracted from the character-level annotations.
Guided by these patterns, the model further revises and supplements the instructions with localized acoustic requirements and assigns grounding labels to the corresponding character-attribute pairs. Only pairs addressed by the final instructions are labeled as grounded. The resulting annotations support both utterance-level instruction following and character-level acoustic grounding. Table~\ref{tab:annotation_statistics} summarizes the annotation statistics.

\subsection{Character-Level Instruction-to-Speech Modeling}
\label{sec:model}

As illustrated in Fig.~\ref{fig:llm}, InstCharVoice models instruction-based speech synthesis as autoregressive generation over character-level speech chunks. 
For each character, the model predicts its grounding label and acoustic attributes before generating the aligned speech tokens. This formulation supports keyword-aware instruction grounding, interpretable acoustic planning, and optional character-level user control.

Let $\mathbf{I}_0$, $\mathbf{C}_0$, and $E_0$ denote the instruction-token sequence, content-token sequence, and speaker embedding, respectively. The speaker embedding is extracted from the reference speech using a pre-trained voiceprint model~\cite{3dspeaker}~\footnote{https://www.modelscope.cn/models/iic/speech\_campplus\_sv\_zh-cn\_3dspeaker\_16k}. The conditioning prefix is constructed as $\mathbf{P}=[\texttt{[SOS]},E_0,\mathbf{I}_0,\mathbf{C}_0,\texttt{[SOT]}]$. 
It is followed by the leading silence tokens $\mathbf{S}_0$ and the ground-truth character chunks. For the $i$-th character, the chunk is represented as $\mathbf{X}^{\mathrm{gt}}_i=[U_i,A_i,\mathbf{S}_i]$, where $U_i$ contains the user-specified acoustic attributes, $A_i$ is the combined embedding of all five acoustic attributes, and $\mathbf{S}_i$ is the speech-token sequence segmented according to the character duration. During training, attributes in $U_i$ are randomly provided or masked, allowing the model to predict the complete five-dimensional attributes from partial user specifications. Thus, the training sequence is $\mathbf{X}^{\mathrm{gt}}=[\mathbf{P},\mathbf{S}_0,\mathbf{X}^{\mathrm{gt}}_1,\ldots,\mathbf{X}^{\mathrm{gt}}_N,\texttt{[EOT]}]$.

    During inference, the model autoregressively predicts the leading silence tokens $\mathbf{\hat{S}}_0$ and character-level speech chunks. 
The speech-token head predicts \texttt{[SOC]} to mark the transition to a new character chunk, while an auxiliary grounding head predicts $\hat{G}_i$ from the same hidden state, indicating whether the character belongs to an instruction-grounded keyword. Grounding prediction provides auxiliary supervision during training and is not fed back into the generation sequence. 
For each character, five attribute-specific heads predict the complete acoustic attributes $\hat{A}_i$ in parallel from a shared hidden state, conditioned on any attributes provided in $U_i$. Unspecified attributes are inferred from the instruction and generation context. The model then autoregressively generates the aligned speech tokens $\mathbf{\hat{S}}_i$ until the next \texttt{[SOC]} or the sequence-ending token \texttt{[EOT]}. 
The generated attributes and speech tokens are fed into WordVoice-FM~\cite{wordvoice} to synthesize the output waveform.

\begin{table*}[t]
\caption{Objective comparison of different systems.}
\label{tab:objective_comparison}
\centering
\resizebox{16cm}{!}{
\begin{tabular}{@{}lccccccc@{}}
\toprule
\textbf{Model}
& \textbf{DNSMOS~$\uparrow$}
& \textbf{CER~$\downarrow$}
& \textbf{kD-MAE~$\downarrow$}
& \textbf{kE-MAE~$\downarrow$}
& \textbf{kP-MAE~$\downarrow$}
& \textbf{kB-RER~$\downarrow$}
& \textbf{kT-RER~$\downarrow$} \\
\midrule
GroundTruth
& 3.616 & --- & 0.0182 & 0.0133 & 0.0094 & 5.26\% & 8.95\% \\
\midrule
FireRedTTS3
& 3.599 & \textbf{2.89\%} & 0.1162 & 0.1409 & 0.2326 & 31.29\% & 47.02\% \\
Qwen3-TTS
& 3.642 & 2.99\% & 0.1057 & 0.1930 & 0.2291 & 28.36\% & 44.56\% \\
CosyVoice3
& 3.583 & 3.12\% & 0.1126 & 0.1411 & 0.2383 & 24.27\% & 47.84\% \\
\midrule
\textbf{InstCharVoice}
& \textbf{3.686} & 3.58\% & \textbf{0.0769} & \textbf{0.1271} & \textbf{0.1857} & \textbf{14.04\%} & \textbf{37.05\%} \\
- w/o Instruction
& 3.560 & 2.91\% & 0.0965 & 0.1431 & 0.2401 & 23.10\% & 44.08\% \\
\bottomrule
\end{tabular}}
\end{table*}

\subsection{Training Objectives}
\label{sec:training}

As illustrated in Fig.~\ref{fig:llm}(c), InstCharVoice-LLM is jointly trained with three cross-entropy objectives: speech-token prediction, character-level grounding prediction, and five-dimensional acoustic attribute prediction. We adopt uncertainty-based loss weighting~\cite{uwloss} to balance these objectives using learnable task uncertainties.

Let $\mathcal{D}$ denote the five acoustic attributes: duration, boundary, energy, pitch, and tone. For classification-based prediction, duration is represented by the number of aligned speech tokens, while energy and pitch are each uniformly quantized into 20 bins over their respective ranges. Boundary and tone use five and seven categories, respectively, following WordVoice-5A~\cite{wordvoice}.
For the $i$-th character and attribute $d\in\mathcal{D}$, the grounding-aware attribute loss is defined as
\begin{align}
\mathcal{L}_{\mathrm{attribute}}
&=\sum_{i=1}^{N}\sum_{d\in\mathcal{D}}
\omega_{i,d}\,
\mathrm{CE}\!\left(\hat{A}_{i,d},A_{i,d}\right),\\
\omega_{i,d}
&=\frac{1}{k_d+1}
\left(
\frac{k_dG_{i,d}}{N_d^{+}}+
\frac{1-G_{i,d}}{N_d^{-}}
\right).
\end{align}
Here, $A_{i,d}$ and $\hat{A}_{i,d}$ denote the ground-truth label and predicted distribution, respectively; $G_{i,d}\in\{0,1\}$ indicates whether the character-attribute pair is grounded by the instruction; and $N_d^{+}$ and $N_d^{-}$ count the grounded and non-grounded pairs for attribute $d$ within the utterance. Attributes with no grounded pairs are trained without grounding-aware reweighting. The factor $k_d$ controls the relative contribution of grounded and non-grounded pairs after group-wise normalization. We set $k_d=5$ for boundary and energy, and $k_d=3$ for the remaining attributes.

We initialize the model from Qwen2.5-0.5B~\cite{qwen2.5} and use the CosyVoice3~\cite{cosyvoice3} tokenizer to extract speech tokens. The model is trained with Adam at a fixed learning rate of $2\times10^{-5}$ on 8 NVIDIA A800 GPUs. 
Training consists of two stages: the model is first trained for five epochs without grounding-aware reweighting, focusing on global instruction-to-speech modeling and character-level acoustic attribute modeling for speech generation; the reweighting scheme is then enabled, and training continues for three additional epochs to further emphasize instruction-relevant acoustic attributes.

\section{Experiments}
\label{sec:prior}

\begin{figure}[t!]
\centering
\includegraphics[width=\linewidth]{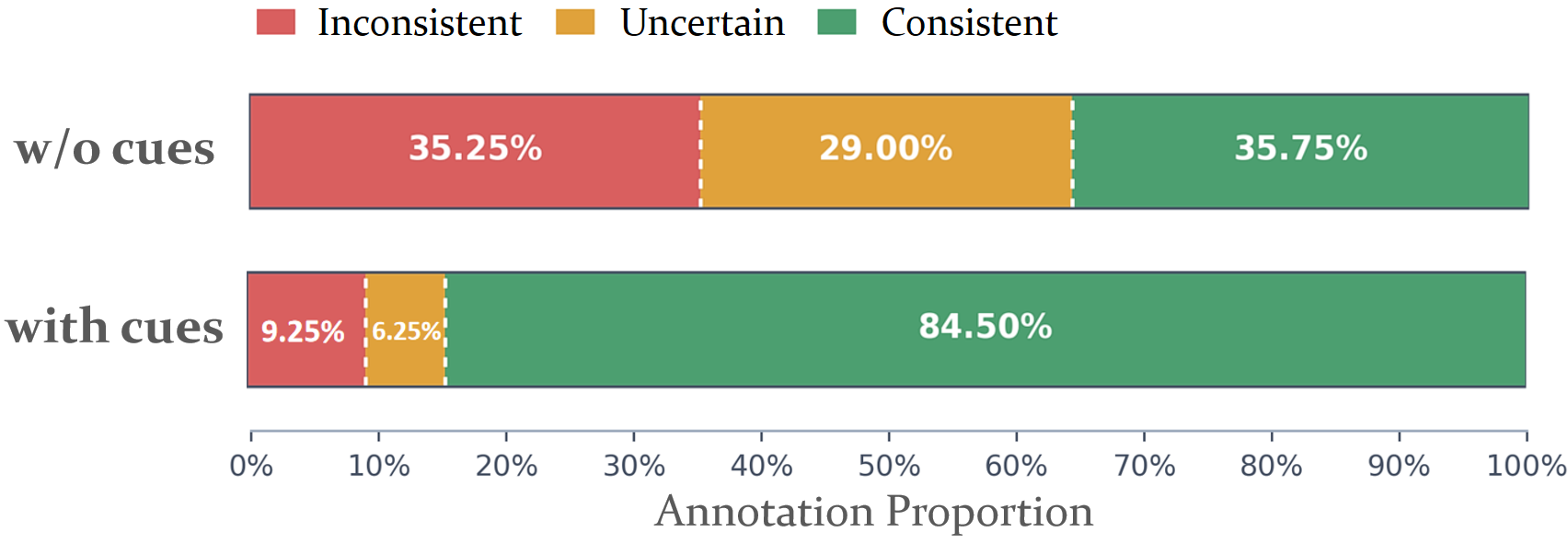}
\caption{Blind evaluation of instruction annotation consistency.}
\label{fig:dataset_annotation}
\end{figure}

\begin{table}[t]
\vspace{-1em}
\centering
\caption{Subjective evaluation results.}
\label{tab:subjective_comparison}
\resizebox{\linewidth}{!}{%
\begin{tabular}{@{}lccc@{}}
\toprule
\textbf{Model} & \textbf{NMOS $\uparrow$} & \textbf{IMOS $\uparrow$} & \textbf{SMOS $\uparrow$} \\ \midrule
FireRedTTS3 & 3.331 ± 0.266 & 3.036 ± 0.306 & \textbf{3.533 ± 0.212} \\
Qwen3-TTS & 3.277 ± 0.276 & 3.357 ± 0.230 & 3.124 ± 0.246 \\
CosyVoice3 & 3.297 ± 0.264 & 3.402 ± 0.253 & 3.374 ± 0.227 \\ \midrule
\textbf{InstCharVoice} & \textbf{3.528 ± 0.219} & \textbf{3.602 ± 0.212} & 3.173 ± 0.260 \\ \bottomrule
\end{tabular}
}
\end{table}

\subsection{Evaluation of Grounded Instruction Annotations}
\label{sec:annotation_evaluation}

We conduct a blind human evaluation to assess instruction-speech consistency and the benefit of character-level acoustic cues. Ten native Chinese speakers each evaluate 40 randomly sampled utterances from WordVoice-5A-zh~\cite{wordvoice}. For each utterance, the same listener evaluates two instruction annotations: \textit{w/o cues}, generated from the speech-text pair in the first annotation stage, and \textit{with cues}, refined using character-level acoustic measurements in the second stage. Listeners judge whether the acoustic characteristics described in each instruction are supported by the speech, using three categories: \textit{inconsistent}, \textit{uncertain}, and \textit{consistent}.

As shown in Fig.~\ref{fig:dataset_annotation}, adding character-level acoustic cues increases consistent judgments from 35.75\% to 84.50\%, while reducing inconsistent judgments from 35.25\% to 9.25\% and uncertain judgments from 29.00\% to 6.25\%. These results indicate that explicit acoustic measurements help the multimodal model produce descriptions that better match the speech, providing a more reliable instruction basis for subsequent local grounding.

\subsection{Main Comparison}
\label{sec:main_comparison}

We evaluate InstCharVoice against three representative ITTS systems: \textbf{FireRedTTS3}~\cite{fireredtts3}, \textbf{Qwen3-TTS}~\cite{qwen3tts}, and \textbf{CosyVoice3}~\cite{cosyvoice3}. We follow the data split of WordVoice-5A and use the same evaluation texts and reference speech across the main comparison, ablation study, and explicit character-level control comparison.

For subjective evaluation, we randomly select 14 instruction-text pairs from the test set and one reference speech sample for each pair. All systems use the same instruction-text pairs and reference speech. Since the selected Qwen3-TTS and FireRedTTS3 models do not support voice cloning in instruction mode, we convert their outputs to the target speaker using Seed-VC~\cite{seedvc}. A total of 27 native listeners rate the synthesized speech on a 5-point scale for naturalness (NMOS), instruction following (IMOS), and speaker similarity (SMOS). We report MOS~\cite{mos} with 95\% confidence intervals. 

For objective evaluation, we use the first 30\% of each ground-truth utterance as reference speech and compare the synthesized speech with the corresponding ground truth. We also include the ground-truth condition and an InstCharVoice variant without instruction input, denoted as \textit{w/o Instruction}. Speech quality and intelligibility are measured using DNSMOS~\cite{dnsmos835} and character error rate (CER) measured by Qwen3-ASR~\cite{qwen3asr}. 
To assess localized acoustic control, we extract character-level attributes using the WordVoice-5A annotation pipeline~\cite{wordvoice} and compare them with the ground-truth annotations. Each metric is computed only on character-attribute pairs grounded by the instruction. We use the prefix ``k'' to denote keyword-level evaluation, followed by D, E, P, B, or T for duration, energy, pitch, boundary, or tone. Mean absolute error (MAE) is computed for duration, energy, and pitch using the normalized values from the original pipeline. Boundary comprises five pause-duration categories, while tone comprises seven categories derived from quadratic fits to within-character pitch contours. For both boundary and tone, we use a relaxed error rate (RER), which treats a prediction within one category of the target as correct. The nonzero acoustic errors in the GroundTruth condition arise from slight differences in timestamp extraction between the evaluation pipeline and the original annotation pipeline.

As shown in Tables~\ref{tab:objective_comparison} and~\ref{tab:subjective_comparison}, InstCharVoice achieves the highest mean NMOS and IMOS, the highest DNSMOS, and the lowest errors across all five keyword-level acoustic metrics among the compared systems. These results demonstrate its strong performance in both speech naturalness and character-level instruction following. In contrast, the other ITTS systems obtain keyword-level acoustic results close to those of the \textit{w/o Instruction} variant, suggesting limited ability to translate natural-language instructions into precise character-level acoustic changes.
InstCharVoice obtains lower SMOS than FireRedTTS3 and CosyVoice3, suggesting a potential trade-off between speaker similarity and fine-grained acoustic control. This comparison is also influenced by the additional Seed-VC conversion applied to Qwen3-TTS and FireRedTTS3. Although its CER is slightly higher, the gap from the best result of 2.89\% is only 0.69 percentage points, indicating that speech intelligibility remains competitive. Overall, InstCharVoice improves naturalness and localized instruction following while largely preserving intelligibility, with speaker similarity remaining an area for improvement.

\begin{table}[t]
\vspace{-1em}
\centering
\caption{Ablation study results.}
\label{tab:ablation}
\resizebox{\linewidth}{!}{%
\begin{tabular}{@{}lccccc@{}}
\toprule
\textbf{Model}
& \textbf{kD-MAE~$\downarrow$}
& \textbf{kE-MAE~$\downarrow$}
& \textbf{kP-MAE~$\downarrow$}
& \textbf{kB-RER~$\downarrow$}
& \textbf{kT-RER~$\downarrow$} \\ 
\midrule
\textbf{Full} & \textbf{0.0769} & \textbf{0.1271} & \textbf{0.1857} & \textbf{14.04\%} & \textbf{37.05\%} \\ 
\midrule
- w/o KP & 0.0805 & 0.1383 & 0.1899 & 16.37\% & 37.82\% \\
- w/o GLW & 0.0859 & 0.1389 & 0.2019 & 20.76\% & 41.82\% \\
- w/o Both & 0.0906 & 0.1583 & 0.2132 & 21.54\% & 42.58\% \\ 
\bottomrule
\end{tabular}
}
\end{table}

\subsection{Ablation Study}
\label{sec:ablation}

We investigate the contributions of keyword prediction (KP) and grounding-aware loss weighting (GLW). The \textit{w/o KP} variant removes the auxiliary keyword prediction head, while \textit{w/o GLW} disables grounding-aware attribute loss weighting. The \textit{w/o Both} variant removes both components while retaining instruction conditioning and five-dimensional acoustic attribute prediction. For each variant, we select the checkpoint with the lowest validation loss .

As shown in Table~\ref{tab:ablation}, removing either component degrades keyword-level acoustic control across all five attributes. Removing GLW causes a larger degradation than removing KP, indicating that grounding-aware loss weighting has a greater impact on localized control. \textit{w/o Both} yields the highest errors across all metrics, supporting the contribution of both components to the full model.

\begin{table}[t]
\vspace{-1em}
\centering
\caption{Character-level control results.}
\label{tab:character_control}
\resizebox{\linewidth}{!}{%
\begin{tabular}{@{}lccccc@{}}
\toprule
\textbf{Model} 
& \textbf{D-MAE~$\downarrow$}
& \textbf{E-MAE~$\downarrow$}
& \textbf{P-MAE~$\downarrow$}
& \textbf{B-ER~$\downarrow$}
& \textbf{T-RER~$\downarrow$} \\ 
\midrule
WordVoice 
& \textbf{0.0348}
& \textbf{0.0484}
& 0.1093
& 12.72\%
& 24.29\% \\
\midrule
\textbf{InstCharVoice} 
& 0.0393
& 0.0544
& \textbf{0.1065}
& \textbf{12.14\%}
& \textbf{21.87\%} \\ 
\bottomrule
\end{tabular}
}
\end{table}

\subsection{Character-Level Control Comparison}

We further compare InstCharVoice with WordVoice under explicit character-level control. Natural-language instructions are removed, and both models receive the same text, reference speech, and complete character-level acoustic attributes. The remaining settings follow the main comparison. Acoustic errors are computed over all characters using the same attribute-specific metrics, except that boundary is evaluated with strict ER rather than RER because its target categories are explicitly provided.

As shown in Table~\ref{tab:character_control}, InstCharVoice achieves overall character-level control performance comparable to WordVoice, with slightly higher errors for duration and energy but lower errors for pitch, boundary, and tone. These differences may reflect trade-offs among the attribute-specific training objectives rather than an overall loss of controllability. The results indicate that incorporating natural-language instruction control largely preserves the original character-level acoustic control capability.

\section{Conclusion}
\label{sec:refs}

We presented \textbf{InstCharVoice}, a unified TTS framework that bridges natural-language instructions and explicit character-level acoustic control. An LLM-assisted annotation pipeline constructs instructions grounded in character-level acoustic evidence, while an autoregressive model predicts instruction-relevant characters and acoustic attributes before generating the corresponding speech tokens. Keyword prediction and grounding-aware loss weighting further improve localized control. Experiments demonstrate improved instruction following and keyword-level acoustic accuracy over representative ITTS systems, together with higher naturalness scores. The framework also retains character-level control performance comparable to WordVoice under explicit attribute conditioning.

The current framework focuses on five measurable acoustic attributes and does not explicitly model abstract factors such as emotion or speaking style. Although localized control is improved, gaps remain in acoustic realization accuracy, intelligibility, and speaker similarity. Evaluation under partial attribute specifications and more complex instructions also remains to be explored. Future work will investigate broader expressive representations and preference-based optimization, including reinforcement learning, to further improve instruction following and perceptual quality.

\begin{small}
\bibliographystyle{IEEEbib}
\bibliography{refs}
\end{small}

\end{document}